\documentclass[12pt,preprint]{aastex}
\usepackage{graphicx}

\newcommand{\sk}[0]{SK }
\newcommand{\Sk}[0]{SK}
\newcommand{\SK}[0]{Super-Kamiokande}
\newcommand{\gf}[0]{``Green's function''}

\shorttitle{GRB Neutrino Search at SK}

\begin{document}

\title{Search for Neutrinos from Gamma-Ray Bursts \\ Using Super-Kamiokande}

\newcounter{foots}
\newcounter{notes}
\newcommand{\authoraticrr}{$^{1}$}
\newcommand{\authoratbu}{$^{2}$}
\newcommand{\authoratbnl}{$^{3}$}
\newcommand{\authoratuci}{$^{4}$}
\newcommand{\authoratcsu}{$^{5}$}
\newcommand{\authoratgmu}{$^{6}$}
\newcommand{\authoratgifu}{$^{7}$}
\newcommand{\authoratuh}{$^{8}$}
\newcommand{\authoratkek}{$^{9}$}
\newcommand{\authoratkobe}{$^{10}$}
\newcommand{\authoratkyoto}{$^{11}$}
\newcommand{\authoratlanl}{$^{12}$}
\newcommand{\authoratlsu}{$^{13}$}
\newcommand{\authoratlsuumd}{$^{13,14}$}
\newcommand{\authoratumd}{$^{14}$}
\newcommand{\authoratmit}{$^{15}$}
\newcommand{\authoratduluth}{$^{16}$}
\newcommand{\authoratsuny}{$^{17}$}
\newcommand{\authoratniigata}{$^{18}$}
\newcommand{\authoratosaka}{$^{19}$}
\newcommand{\authoratseoul}{$^{20}$}
\newcommand{\authoratshizuokasc}{$^{21}$}
\newcommand{\authoratshizuoka}{$^{22}$}
\newcommand{\authorattohoku}{$^{23}$}
\newcommand{\authorattokyo}{$^{24}$}
\newcommand{\authorattokai}{$^{25}$}
\newcommand{\authorattit}{$^{26}$}
\newcommand{\authoratwarsaw}{$^{27}$}
\newcommand{\authoratuw}{$^{28}$}

\newcommand{\addressoficrr}[1]{$^{1}$ #1 }
\newcommand{\addressofbu}[1]{$^{2}$ #1 }
\newcommand{\addressofbnl}[1]{$^{3}$ #1 }
\newcommand{\addressofuci}[1]{$^{4}$ #1 }
\newcommand{\addressofcsu}[1]{$^{5}$ #1 }
\newcommand{\addressofgmu}[1]{$^{6}$ #1 }
\newcommand{\addressofgifu}[1]{$^{7}$ #1 }
\newcommand{\addressofuh}[1]{$^{8}$ #1 }
\newcommand{\addressofkek}[1]{$^{9}$ #1 }
\newcommand{\addressofkobe}[1]{$^{10}$ #1 }
\newcommand{\addressofkyoto}[1]{$^{11}$ #1 }
\newcommand{\addressoflanl}[1]{$^{12}$ #1 }
\newcommand{\addressoflsu}[1]{$^{13}$ #1 }
\newcommand{\addressofumd}[1]{$^{14}$ #1 }
\newcommand{\addressofmit}[1]{$^{15}$ #1 }
\newcommand{\addressofduluth}[1]{$^{16}$ #1 }
\newcommand{\addressofsuny}[1]{$^{17}$ #1 }
\newcommand{\addressofniigata}[1]{$^{18}$ #1 }
\newcommand{\addressofosaka}[1]{$^{19}$ #1 }
\newcommand{\addressofseoul}[1]{$^{20}$ #1 }
\newcommand{\addressofshizuokasc}[1]{$^{21}$ #1 }
\newcommand{\addressofshizuoka}[1]{$^{22}$ #1 }
\newcommand{\addressoftohoku}[1]{$^{23}$ #1 }
\newcommand{\addressoftokyo}[1]{$^{24}$ #1 }
\newcommand{\addressoftokai}[1]{$^{25}$ #1 }
\newcommand{\addressoftit}[1]{$^{26}$ #1 }
\newcommand{\addressofwarsaw}[1]{$^{27}$ #1 }
\newcommand{\addressofuw}[1]{$^{28}$ #1 }

\author{
{\bf The Super-Kamiokande Collaboration} \\
\vspace{0.2cm}
%
S.~Fukuda\authoraticrr,
Y.~Fukuda\authoraticrr,
M.~Ishitsuka\authoraticrr, 
Y.~Itow\authoraticrr,
T.~Kajita\authoraticrr, 
J.~Kameda\authoraticrr, 
K.~Kaneyuki\authoraticrr,
K.~Kobayashi\authoraticrr, 
Y.~Koshio\authoraticrr, 
M.~Miura\authoraticrr
% ...
%
A.L.~Stachyra\authoratuw, 
R.J.~Wilkes\authoratuw 
%
%
%
\\
\smallskip
\footnotesize
\it
\addressoficrr{Institute for Cosmic Ray Research, University of Tokyo, Kashiwa,Chiba 277-8582, Japan}\\
\addressofbu{Department of Physics, Boston University, Boston, MA 02215, USA}\\
\addressofbnl{Physics Department, Brookhaven National Laboratory, Upton, NY 11973, USA}\\
\addressofuci{Department of Physics and Astronomy, University of California, Irvine, Irvine, CA 92697-4575, USA }\\
\addressofcsu{Department of Physics, California State University, Dominguez Hills, Carson, CA 90747, USA}\\
\addressofgmu{Department of Physics, George Mason University, Fairfax, VA 22030, USA }\\
\addressofgifu{Department of Physics, Gifu University, Gifu, Gifu 501-1193, Japan}\\
\addressofuh{Department of Physics and Astronomy, University of Hawaii, Honolulu, HI 96822, USA}\\
\addressofkek{Institute of Particle and Nuclear Studies, High Energy Accelerator Research Organization (KEK), Tsukuba, Ibaraki 305-0801, Japan }\\
\addressofkobe{Department of Physics, Kobe University, Kobe, Hyogo 657-8501, Japan}\\
\addressofkyoto{Department of Physics, Kyoto University, Kyoto 606-8502, Japan}\\
\addressoflanl{Physics Division, P-23, Los Alamos National Laboratory, Los Alamos, NM 87544, USA }\\
\addressoflsu{Department of Physics and Astronomy, Louisiana State University, Baton Rouge, LA 70803, USA }\\
\addressofumd{Department of Physics, University of Maryland, College Park, MD 20742, USA }\\
\addressofmit{Department of Physics, Massachusetts Institute of Technology, Cambridge, MA 02139, USA}\\
\addressofduluth{Department of Physics, University of Minnesota Duluth, MN 55812-2496, USA}\\
\addressofsuny{Department of Physics and Astronomy, State University of New York, Stony Brook, NY 11794, USA}\\
\addressofniigata{Department of Physics, Niigata University, Niigata, Niigata 950-2181, Japan }\\
\addressofosaka{Department of Physics, Osaka University, Toyonaka, Osaka 560-0043, Japan}\\
\addressofseoul{Department of Physics, Seoul National University, Seoul 151-742, Korea}\\
\addressofshizuokasc{International and Cultural Studies, Shizuoka Seika College, Yaizu, Shizuoka, 425-8611, Japan}\\
\addressofshizuoka{Department of Systems Engineering, Shizuoka University, Hamamatsu, Shizuoka 432-8561, Japan}\\
\addressoftohoku{Research Center for Neutrino Science, Tohoku University, Sendai, Miyagi 980-8578, Japan}\\
\addressoftokyo{The University of Tokyo, Tokyo 113-0033, Japan }\\
\addressoftokai{Department of Physics, Tokai University, Hiratsuka, Kanagawa 259-1292, Japan}\\
\addressoftit{Department of Physics, Tokyo Institute for Technology, Meguro, Tokyo 152-8551, Japan }\\
\addressofwarsaw{Institute of Experimental Physics, Warsaw University, 00-681 Warsaw, Poland }\\
\addressofuw{Department of Physics, University of Washington, Seattle, WA 98195-1560, USA}\\
}

\newcommand{\presaddressfermi}{Present address: Enrico Fermi Institute, University of Chicago, Chicago, IL 60637}
\newcommand{\presaddressmit}{Present address: Department of Physics, Massachusetts Institute of Technology, Cambridge, MA 02139, USA}
\newcommand{\presaddressutah}{Present address: Department of Physics, University of Utah, Salt Lake City, UT 84112, USA}
\newcommand{\presaddresskorea}{Present address: Korea Research Institute of Standards and Science, Yusong P.O. Box 102, Taejon, 305-600, Korea}

\newpage

\begin{abstract}

Using the \SK\, neutrino observatory, a search was conducted for neutrinos produced in coincidence with gamma-ray bursts observed by the BATSE detector.  \SK\, data in the neutrino energy range of $\rm7\,MeV\sim100\,TeV$ were analyzed.  For gamma-ray bursts that occurred between 1996 April and 2000 May, no statistically significant signal in excess of the background levels was detected.  Implied upper limits on associated GRB neutrino production are presented.

\end{abstract}


\section{Introduction}

Gamma-Ray Bursts (GRBs) are some of the most luminous astrophysical objects observed, and little is known about them.  They have been observed to release an enormous amount of energy in the form of gamma-rays, and models predict that a considerable portion of the total energy may be carried away by neutrinos created during the burst.  Unlike the GRB photons, which may scatter or be absorbed before they reach the Earth, the neutrinos, due to their small interaction cross section, arrive at the Earth virtually unaffected.  Thus, studying neutrinos from GRBs can provide insight in understanding the processes that underlie these mysterious phenomena.  This paper presents the results of a time correlation analysis between GRBs and \SK\,(\Sk) events from the neutrino data samples used in the solar neutrino and atmospheric neutrino analyses, as well as a direction-time correlation analysis using the ``upward-going'' muon data sample.

\sk is a 50 kton water Cherenkov detector located in the Kamioka Mine in Gifu, Japan.  The cylindrical detector is divided into an inner and outer detector (ID and OD, respectively) by a stainless-steel frame structure that serves as an optical barrier and a mounting point for all photo-multiplier tubes (PMTs).  Cherenkov light in the ID is collected by 11,146 inward-facing $50\,$cm  PMTs mounted uniformly on the wall, providing $40\%$ photo-cathode coverage.  In the OD, 1885 outward-facing $20\,$cm PMTs monitor the $2.5\,$m thick veto region.  The veto is used to tag incoming particles and serves as a passive shield for gamma activity from the surrounding rock.  Detailed descriptions of the \sk detector can be found elsewhere \cite{r:solar,r:nakahata}.

The most widely accepted theoretical description for gamma-ray production by GRBs is the relativistic fireball shock model \cite{r:fireball}, in which GRBs are produced when relativistic ejecta from a ``central engine'' are slowed down by interactions, either with an external medium (the external shock model) or among different layers within the ejecta themselves (the internal shock model) \cite{r:narayan}.  The strongest confirmation of the fireball model comes from observations of the GRB afterglows \cite{r:katz-piran,r:vietri_AG,r:sari}.  Neutrino production in this model may be due primarily to $p-\gamma$ interactions ($\rm E_\nu\sim10^{14}-10^{19}\,eV$ or higher) \cite{r:waxman_HE,r:depaolis,r:guetta,r:wax-bahc,r:vietri} or $p-n$ collisions ($\rm E_\nu\sim10\,MeV-10\,GeV$) \cite{r:waxman_HE,r:depaolis,r:bahc-mesz,r:mesz-rees,r:kumar}.  Great uncertainty still surrounds the ``central engine'' of the GRB (i.e. the progenitor of the fireball).  The current scenarios include a collapse of massive objects (hypernovae, supernovae, supranovae, collapsars) and mergers of binary systems (black holes, neutron stars, white dwarfs, helium stars).  Almost all models involve accretion disks and jet formation, which gives rise to relativistic fireballs due to extremely high temperatures associated with accretion.  A review of these many models can be found in M\'{e}sz\'{a}ros (2001) and Piran (2000).  

Because of the lack of specificity in models, the present analysis is spectrum-independent and was done for all neutrino energies to which \sk is sensitive ($\rm7\,MeV\sim100\,TeV$).  The low energy (LE) neutrino sample consists of recoil electron events from $\nu\,$-$\,e$ elastic scattering of solar neutrinos, as well as background events due to radioactivity and interactions and decays of cosmic-ray muons.  The neutrino energies in the LE sample are in the range: $E_\nu=7\sim80\,$MeV.  More details on the LE data can be found elsewhere \cite{r:solar}.  The high energy (HE) data sample contains fully-contained (all products of the neutrino interaction stop in the ID) and partially-contained (some products of the neutrino interaction penetrate into the OD) electron and muon events from neutrino-nucleon interactions of atmospheric neutrinos \cite{r:atm}.  The neutrino energies in the HE sample are in the range $E_\nu\rm=200\,MeV\sim200\,GeV$.  The upward-going muon (upmu) sample yields the highest energy events at \Sk, consisting of upward-going muons created by neutrino interactions in the rock beneath the detector \cite{r:SKupthru,r:SKupstop}.  Typical parent neutrino energies in the upmu sample are $10\,$GeV and $100\,$GeV for muons that stop in the ID and the muons that penetrate into the OD, respectively, and they span the range $E_\nu\rm=2\,GeV\sim100\,TeV$.  Both HE and upmu sample events are predominantly due to atmospheric neutrino interactions.

\section{GRB-neutrino correlation analysis}

In order to search for a possible GRB neutrino signal, we conducted a time correlation analysis of GRBs using \Sk's LE and HE events, and a direction-time correlation analysis using \Sk's upmu events.  The list of GRBs selected for the analysis was obtained from the BATSE online catalog \cite{r:batse} and the non-triggered supplement\footnote{
These include bursts which did not activate the real-time burst detection system onboard the BATSE spacecraft and were discovered during an off-line search.}
to the BATSE catalog \cite{r:kommers}.  BATSE (Burst And Transient Source Experiment) was a high energy astrophysics experiment in Earth orbit on NASA's Compton Gamma-Ray Observatory.  From the official start of data taking at \sk on 1996 April 1 until the end of the BATSE mission in 2000 May, a total of 1454 GRBs were selected to match the four years of operational coincidence of the two experiments.  A total of 1371 GRBs were used in the LE and HE correlation analysis, because the LE and HE data began being collected two months later (1996 May 31).

The goal of all three (LE, HE, and upmu) correlation analyses was to search for an excess in the number of events correlated with GRBs above expected background.  All events in both the LE and HE samples were considered background for this analysis, since a possible GRB signal would provide a negligible contribution to the sample.  The LE and HE mean background rates were assumed to be constant and were measured to be $(79.8\pm0.3)\times10^{-5}\,\rm s^{-1}$ and $(9.9\pm0.1)\times10^{-5}\,\rm s^{-1}$, respectively.  In both cases no statistically significant time variation of the background rate was found.  The background for the upmu sample was calculated by a Monte-Carlo simulation for each GRB separately, because there is directional variation of the events.

The three analyses were naturally separated into two sections.  The LE and HE {\it time} correlation analyses are presented together because their background rates can be calculated from the data, and they both use the official fiducial volume of the detector for data collection.  Since it is not always possible to infer the direction of the parent neutrino from the direction of the event, only a time correlation analysis was nominally performed, and a directional correlation was checked for the time-correlated events.  In contrast, in the upmu sample the background rates were calculated by a simulation, and a much bigger fiducial volume (including the surrounding rock) was used.  Also, because of a strong directional correlation between the parent neutrino and the resulting muon, a {\it direction-time} correlation analysis was possible for the upmu data sample.

\subsection{GRB Search with LE and HE Neutrinos}
\label{s:lehe}

Since most GRB models predict neutrinos in coincidence with the photons, and there is no appreciable GRB neutrino flight-time delay with respect to the photons (assuming $m_\nu$$<$$0.01\,$eV/c$^2$, $E_\nu$$>$$7\,$MeV, $z$$\sim$$1$), a $\pm10\,$s window centered on the GRB time was first used in the analysis.  This tight window would detect possible GRB neutrinos arriving virtually simultaneously with the onset of the gamma-ray burst.  Next, a time correlation window of $\pm100\,$s centered on the GRB time was used because $90\%$ of the GRBs analyzed have T90$\leq$$100\,$s(\footnote{
T90 is the duration (measured by BATSE) required for integration of 90\% of the burst fluence for each gamma-ray burst.}).
A conservative $\pm1000\,$s window was also used because virtually all analyzed GRBs ($>$$99.9\%$) have T90$\leq$$1000\,$s.  In each case the time window was taken as $\pm$ in order to detect neutrinos that may precede or lag the photons due to some unknown GRB process.  Finally, a search window of $1\,$hr was used to scan the $24\,$hr period before each GRB (24 independent searches) in order to search for neutrinos that may have been produced in the GRB from supernova-like (SN-like) processes, and would precede the photons.  Similarly, a $1\,$hr window was also used to scan the $24\,$hr period after each GRB in an attempt to detect any neutrinos that may have been created in a GRB afterglow (AG) process.

\begin{figure}[h]
\epsscale{0.75}
\plotone{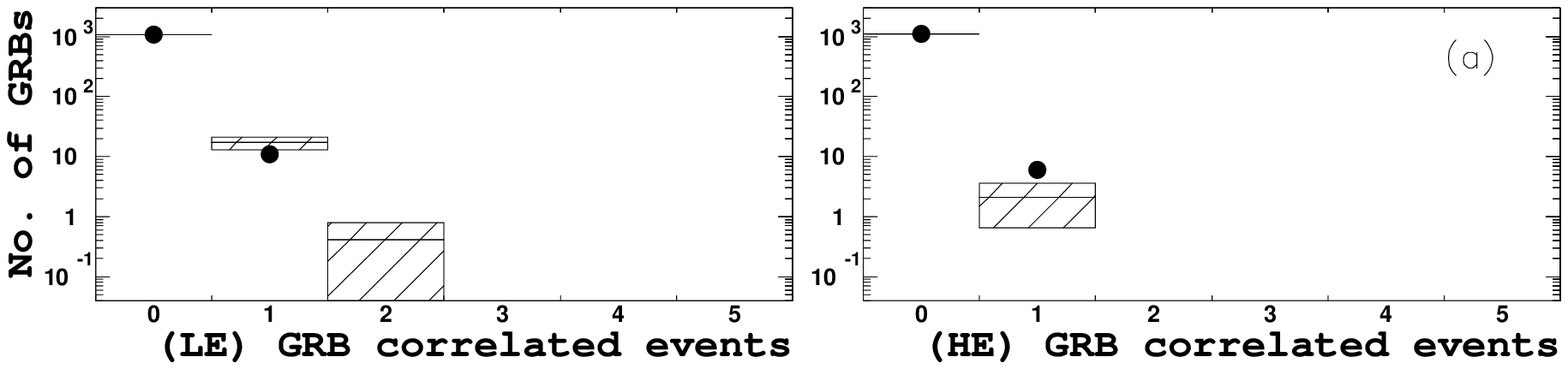}\vspace{1.0em}
\plotone{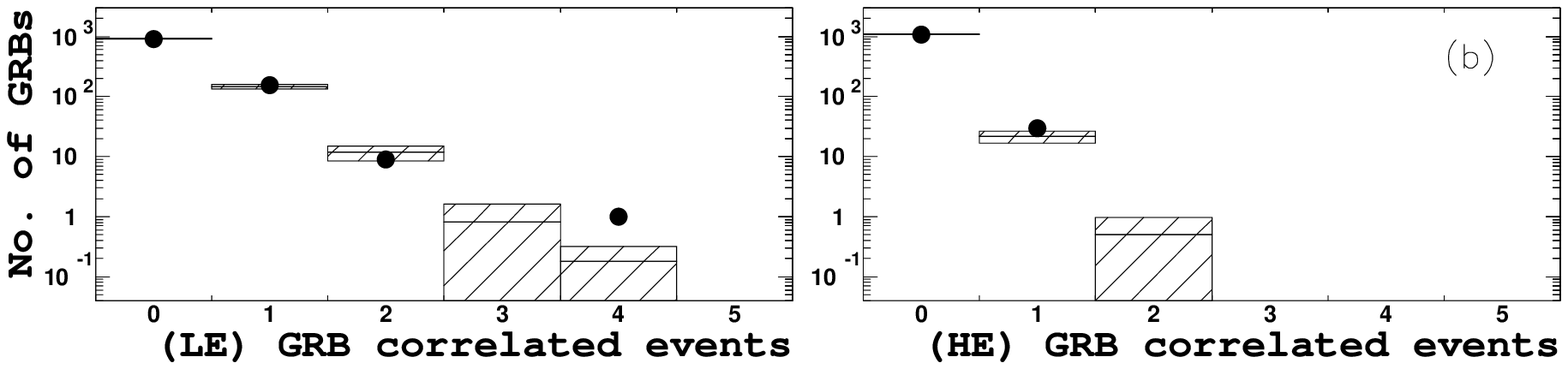}\vspace{1.0em}
\plotone{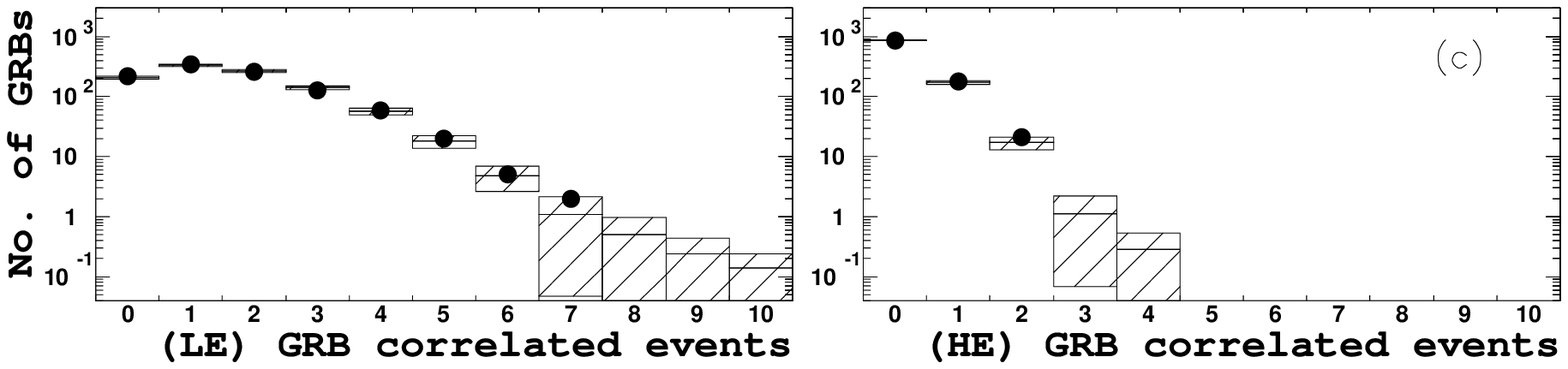}\vspace{1.0em}
\plotone{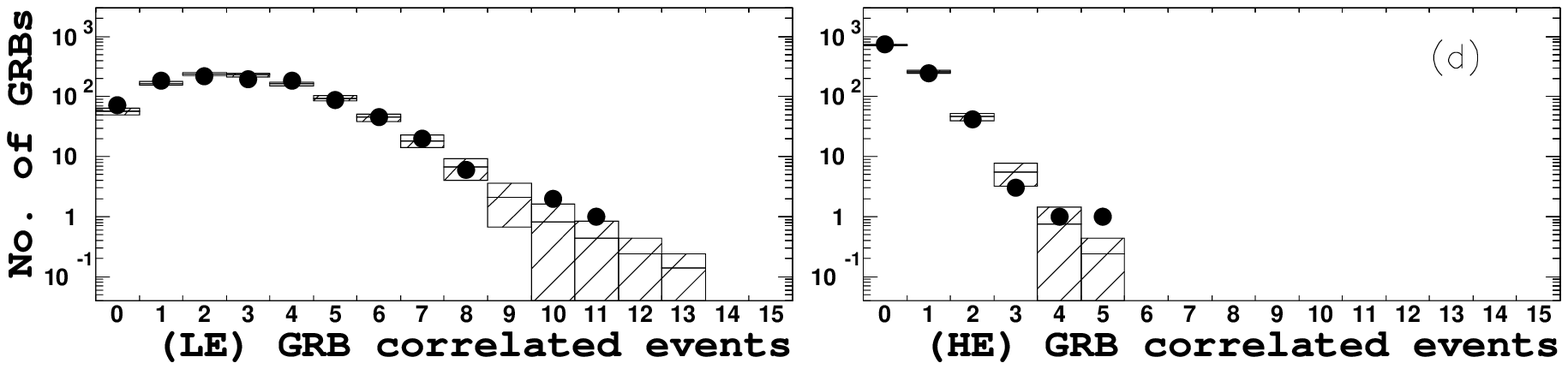}\vspace{1.0em}
\plotone{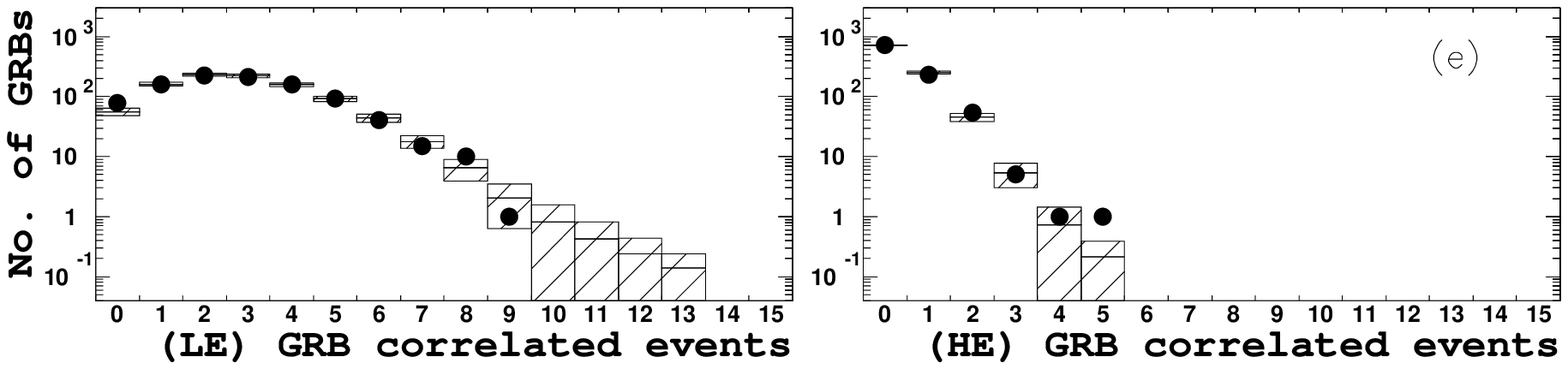}
\caption{\label{f:time} \small Distributions of the number of GRB neutrino signal candidate events ({\it points}) with the expected Poisson fluctuation of the background for the $\pm10\,$s (a), $\pm100\,$s (b), and $\pm1000\,$s (c) time correlation windows as well as the $1\,$hr window between 8 and 7$\,$hr (d) {\it prior} to GRB time (supernova-like process) and the $1\,$hr window between 4 and 5$\,$hr (e) {\it after} the GRB time (GRB afterglow), respectively.  The shaded region of the background prediction in each bin represents the $1\,\sigma$ statistical error bars.  For each time window, the left plot is the result for the LE sample, and the right plot for the HE sample.}
\end{figure}

Figure~\ref{f:time} shows the distribution of the number of observed signal candidate events for each time correlation window, compared with the expectation from Poisson fluctuations of the measured background for the LE (left) and the HE (right) data samples.  For the SN-like and AG searches, only the $1\,$hr window with the largest deviation of the signal from the background is shown (out of 24 windows searched in each case).  The shaded region of the background prediction in each bin represents the $1\,\sigma$ statistical error bars.  As demonstrated by Figure~\ref{f:time}, the observed distribution of events is in good agreement with expectations from the background.  

Table~\ref{t:results} summarizes the bin with the largest number of observed events for each time correlation window.  It shows the number of GRBs in the analysis sample, the expected background for a single GRB, the maximum number of observed events, the number of GRB candidates with that number of events (in cases of one or two GRBs, their BATSE catalog number is given), and the probability to get at least that many GRB candidates, given the expected background and the total number of GRBs.  The top entry of each search window corresponds to the LE analysis and the bottom entry to the HE analysis.  The last entry in the table represents the result from the upmu search analysis (see \S\ref{s:upmu}).  The number of GRBs is different in each search, because the three data samples employ different live-time calculation methods.  Moreover, since we require that the detector live-time around each GRB at least equals the length of the time search window, longer search windows yielded a smaller number of GRBs to be analyzed.  In both the SN-like and AG searches the probabilities include a trials factor of 24 to account for the 24 search windows used.  All the probabilities are consistent with the expectation from random fluctuations of the background.

\begin{figure}[h]
\epsscale{0.75}
\plotone{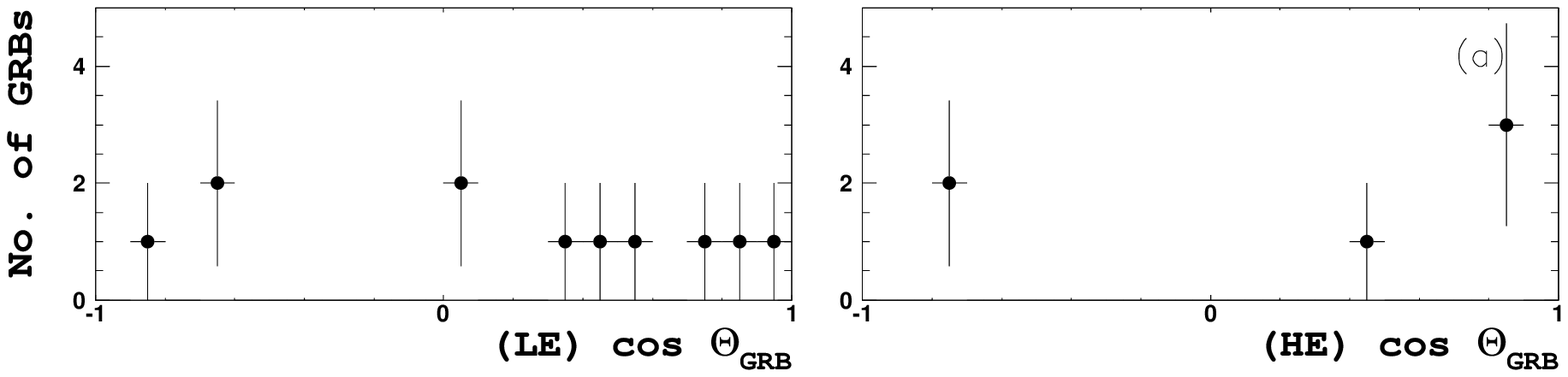}\vspace{1.0em}
\plotone{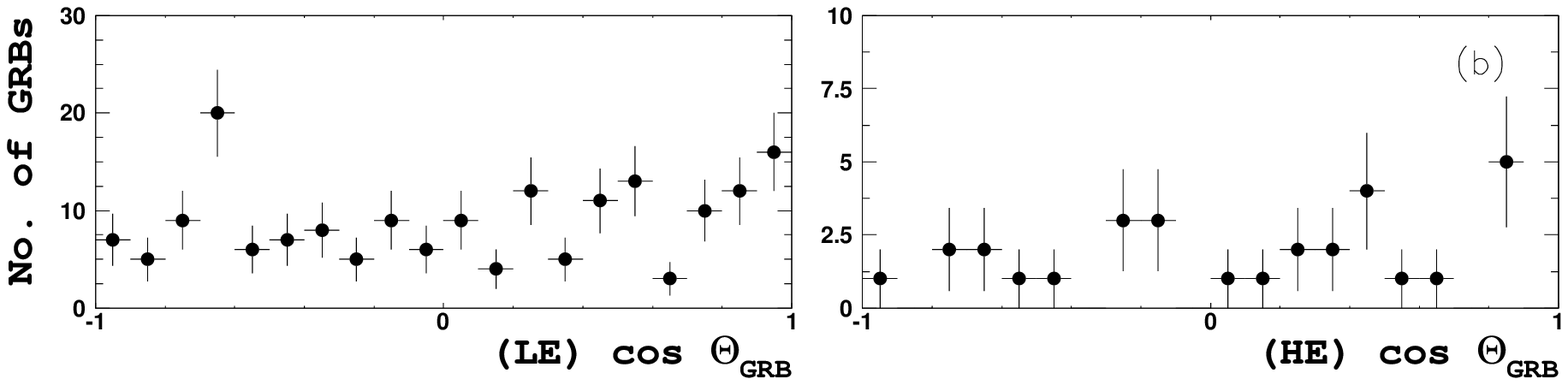}\vspace{1.0em}
\plotone{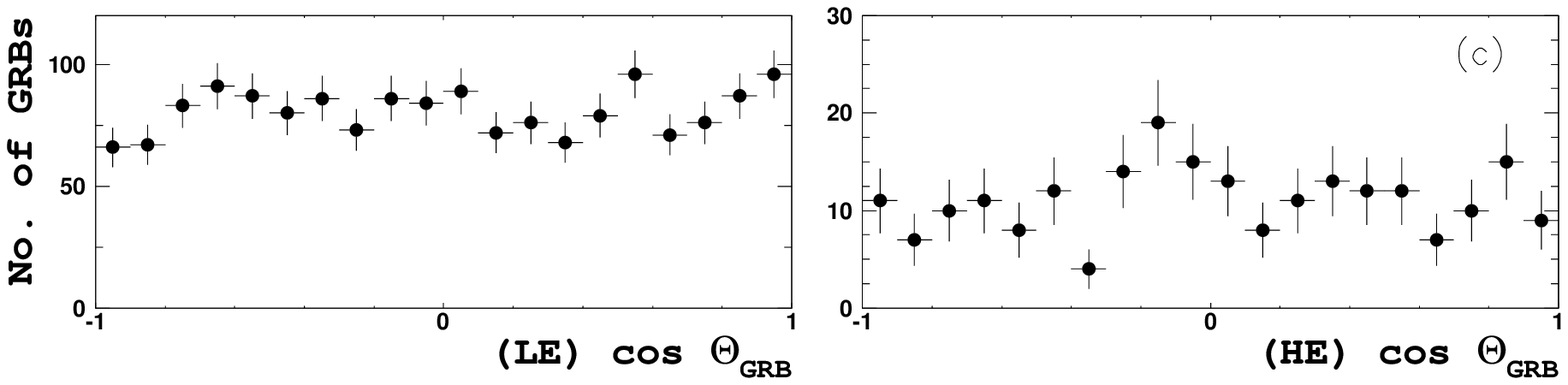}\vspace{1.0em}
\plotone{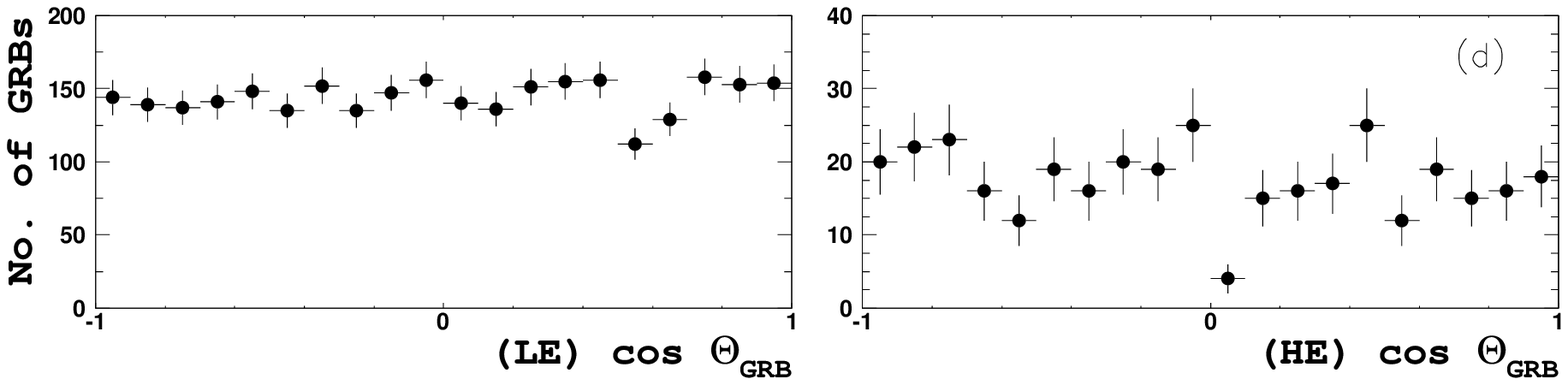}\vspace{1.0em}
\plotone{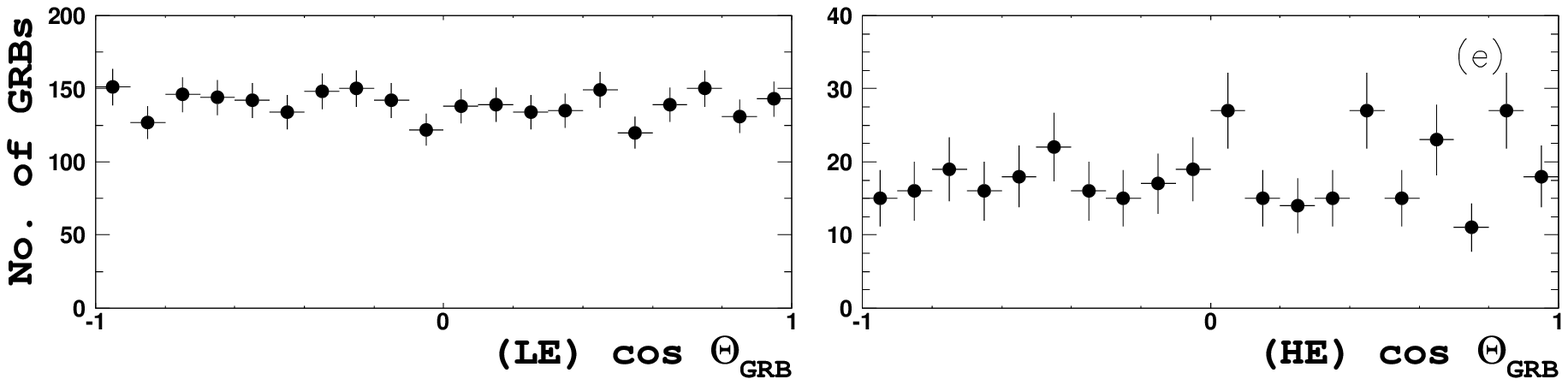}
\caption{\label{f:cos} \small Distributions of the directional correlation between GRBs and the neutrino signal candidate events for the $\pm10\,$s (a), $\pm100\,$s (b), and $\pm1000\,$s (c) time correlation windows as well as the $1\,$hr window between 8 and 7$\,$hr (d) {\it prior} to GRB time (supernova-like process) and the $1\,$hr window between 4 and 5$\,$hr (e) {\it after} the GRB time (GRB afterglow), respectively.  For each time window, the left plot is the result for the LE sample, and the right plot for the HE sample.}
\end{figure}

For some neutrino interactions, a correlation is expected between the parent neutrino direction and the direction of the interaction product (eg. lepton).  In order to further examine whether the observed correlated events are possibly signal rather than background, a directional correlation check was performed for the GRBs with at least one LE or HE event in any of their search time windows.  For each GRB-event correlated pair, the cosine of the angle between their directions (cos $\Theta_{\rm GRB}$) was calculated.  Figure~\ref{f:cos} shows the distributions of cos $\Theta_{\rm GRB}$ for each time correlation window.  A possible directional correlation signal would be manifested as a rise of the distribution at cos $\Theta_{\rm GRB}$ = 1.  All the distributions are flat within the background fluctuations, which demonstrates no directional correlation.

\subsection{GRB Search with Upward-Going Muons}
\label{s:upmu}

The expected background for each GRB was, as stated earlier, calculated by simulating the atmospheric neutrino induced upward-going muon events with a 40-year equivalent Monte-Carlo.  These Monte-Carlo simulations use Bartol atmospheric neutrino flux~\cite{r:Bartol}, the GRV94 parton distribution function~\cite{r:GRV94}, and energy loss mechanisms of muons in the rock \cite{r:Lipari}.  In the simulation we added suppression of $\nu_\mu$ flux due to neutrino oscillations with the oscillation parameters given by $\sin^2 2{\theta} = 1$ and $\Delta \mbox{m}^2 =2.5 \times 10^{-3} \mbox{eV}^2$ \cite{r:bestfit}, which are \Sk's best-fit $\nu_\mu\rightarrow\nu_{\tau,x}$ oscillation parameters for the data sample used. 

The direction of an upmu is strongly correlated with the incident
neutrino direction. The RMS angle between a muon and its parent neutrino
is $3.7^\circ$, assuming an atmospheric neutrino spectrum (harder GRB neutrino
spectra would result in a smaller angular separation of the muons and
GRBs). Taking into account this separation and the BATSE directional 
uncertainty for each GRB (average of $\sim5^\circ$), we chose a direction search cone of a $15^\circ$ half angle between the GRB and the muon.


Because we use a 15$^\circ$ direction cut, in this analysis we only used the conservative $\pm1000\,$s time window to perform the search.  We found only one upmu which was correlated in time (within $\pm1000\,$s) and direction (within 15$^\circ$) with a GRB.  Since there was only one correlation found, it is only presented as the last entry in Table~\ref{t:results} (see \S\ref{s:lehe} for the description of the table).  Again, the probability is consistent with the expectation from the background.  

In addition, we searched for upward-going muons that were correlated in direction with a GRB ($\leq$$15^{\circ}$), and correlated in time with each other ($\leq$$1000\,$s), all within a $\pm24\,$hr window of a single GRB, again in hope to detect any possible SN-like or AG neutrinos.  No such clusters of muons were found.

\section{Upper Limits on Neutrino Fluence}
\label{s:flux}

In the absence of any clear neutrino signal from the GRBs, $90\%$ C.L. GRB neutrino and antineutrino fluence upper limits were calculated.  In principle, a fluence limit, $F\rm[cm^{-2}$] is given by
\begin{equation}\label{e:f}
F = \frac{N_{90}}{N_T\int{\sigma(E_\nu)\epsilon(E_\nu)\lambda(E_\nu)dE_\nu}}\;,
\end{equation}
where $N_{90}$ is the 90\% C.L. limit on the total number of neutrino interactions detected in \Sk, $N_T$ is the number of interaction targets (electrons or nucleons), $\sigma(E_\nu)$ is the total neutrino cross section as a function of neutrino energy, $\epsilon(E_\nu)$ is the detector efficiency as a function of neutrino energy, and $\lambda(E_\nu)$ is the neutrino energy spectrum normalized to unity.  

Without assuming the neutrino spectrum (there is no generally accepted GRB neutrino spectrum), it is impossible to directly calculate the neutrino fluence.  A quantity we can calculate without assuming a spectrum is a fluence limit \gf, $\Phi(E_\nu)$, which is obtained by replacing $\lambda(E_\nu)$ by a delta function $\delta(E-E_\nu)$:
\begin{equation}\label{e:phi}
\Phi_{_{\rm LE,HE}}(E_\nu) = \frac{N_{90}}{N_T \int{\sigma(E_\nu')\epsilon(E_\nu')\delta(E_\nu-E_\nu')dE_\nu'}}\;,
\end{equation}
or in case of the upmu analysis:
\begin{equation}\label{e:phi2}
\Phi_{\rm upmu}(E_\nu) = \frac{N_{90}}{AN_a \int\left[\int_{E_{th}}^{E_\nu'}\frac{d\sigma(E_\nu')}{dE_\mu}r(E_\mu)dE_\mu\right]\delta(E_\nu-E_\nu')dE_\nu'}\;,
\end{equation}
where $A$ is the average effective area for neutrino detection, $r(E_\mu)$ is the effective range of the muon (including the rock), and $N_a$ is Avogadro's number.  For the upmu sample, $\Phi(E_\nu)$ is calculated slightly differently, because the fiducial volume includes the rock surrounding the detector, which also results in much lower fluence upper limits.

The fluence limit \gf\, represents the fluence limit on monoenergetic neutrinos at different specific energies.  In order to obtain the total integrated neutrino fluence limit, $F$, one needs to convolute $\Phi(E_\nu)$ with a particular neutrino spectrum, $\lambda(E_\nu)$:
\begin{equation}\label{e:tot}
F = \left[\int{\frac{\lambda(E_\nu)}{\Phi(E_\nu)}dE_\nu}\right]^{-1} \;.
\end{equation}

The 90\% C.L. upper limit on the number of \sk events per GRB, $N_{90}$, was obtained by assuming a Poisson process with background \cite{r:limit}.  Table \ref{t:number} shows this number limit used in the calculation of $\Phi(E_\nu)$ for all three data samples.  The background value in the table represents the total cumulative number of expected background events for all the GRBs in the analysis for the given search window; the signal value is the total, cumulative number of events detected in that search window for all GRBs.  In order to set a limit on neutrino fluence per GRB, a total number limit was first calculated from these combined values, and then divided by the number of GRBs in the search.  For the LE and HE samples, the $\pm100\,$s window was a natural choice for calculating the upper limits, because for 90\% of the GRBs, $100\,$s is the time required for integrating 90\% of the gamma-ray signal.  Since in the HE sample we can distinguish e-like and $\mu$-like events, the $N_{90}$ upper limits are given for e-like ($\nu_e$, $\bar{\nu}_e$) and $\mu$-like ($\nu_\mu$, $\bar{\nu}_\mu$) separately.  In this way we gain about 35\% in sensitivity compared to a result from combined HE data.  For the upmu sample only one search window was used.

The fluence limits presented here reflect the sensitivity of \Sk's solar neutrino, atmospheric neutrino, and upward-going muon official data samples to possible GRB neutrinos.  The detection cross section for $\nu_e$, $\nu_\mu$, and $\bar{\nu}_\mu$ at energies below $100\,$MeV (solar neutrino sample) comes predominantly from neutrino-electron elastic scattering ($\nu\;e^- \rightarrow \nu\;e^-$); the cross section for $\bar{\nu}_e$ is dominated by inverse beta decay scattering ($\bar{\nu}_e\;p \rightarrow n\;e^+$).  All four types of neutrinos at energies above $100\,$MeV (atmosperic neutrino sample and upward-going muon sample) are detected at \sk by neutrino-nucleon charged current quasi-elastic scattering ($\nu\;N \rightarrow l\;N'$) and deep inelastic scattering (single, multiple, and coherent $\pi$ production).  In the upmu sample, however, only the interactions that produce $\mu^\pm$ are considered, and therefore, we are limited to calculating the fluence limit only for $\nu_\mu$ and $\bar{\nu}_\mu$.

The efficiency of \sk at each neutrino energy was calculated by generating MC interactions of monoenergetic neutrinos and simulating the detector response to the resulting interaction products.  The simulation was based on the \sk solar neutrino analysis (for LE sample), atmospheric neutrino analysis (for HE sample), and upward-going muon analysis (for upmu sample).  The fluence limit obtained from the HE sample was calculated only up to $200\,$GeV ($40\,$GeV for $\nu_e$), because the understanding of the detector response is limited at these high energies.  For the lowest energies of the HE analysis, the fluence limit for $\nu_\mu$ and $\bar{\nu}_\mu$ are sllightly worse than expected, because of low detector efficiency and a lower scattering cross section at the lowest energies of the sample.

\begin{figure}[h]
\epsscale{0.5}
\plotone{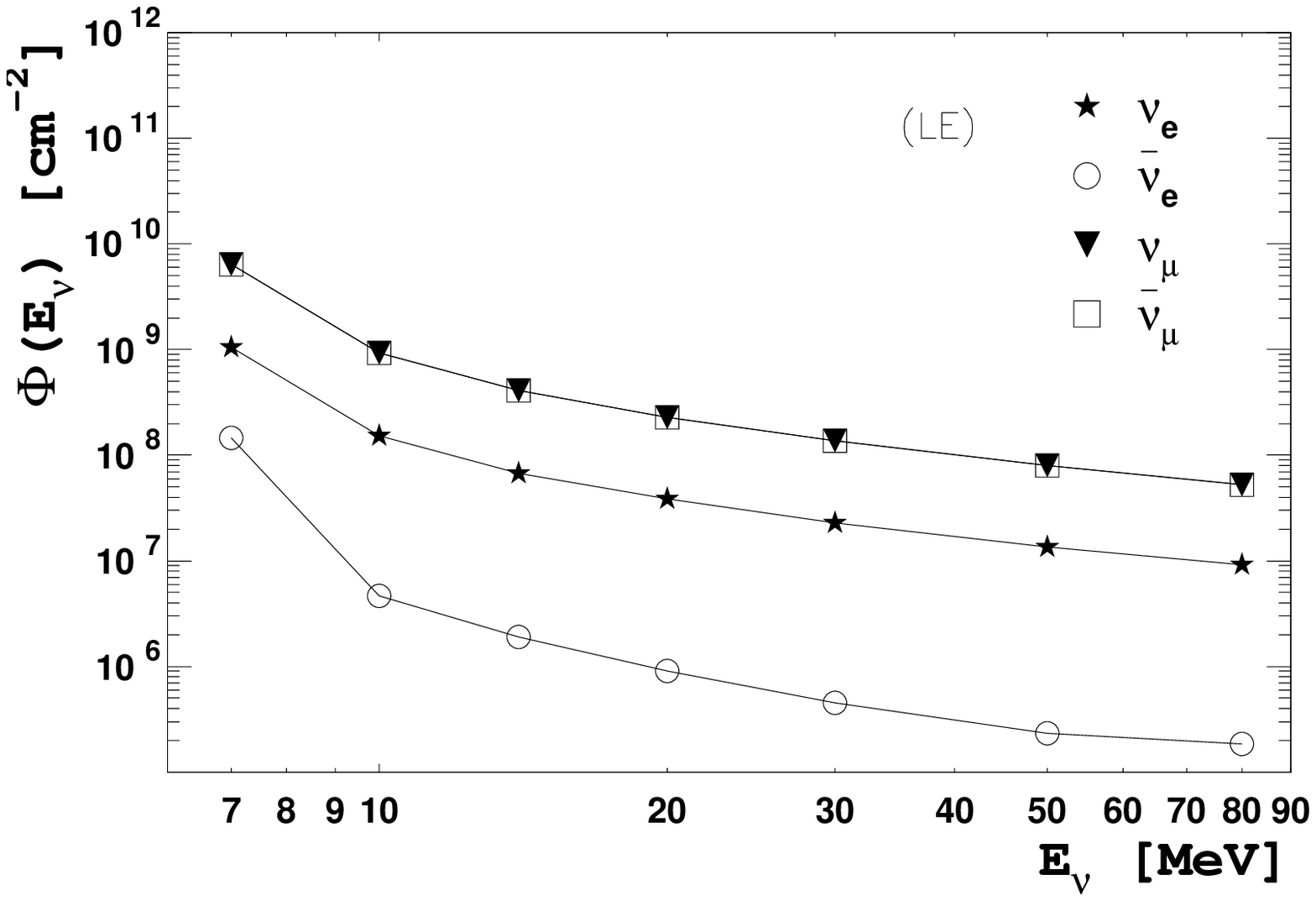}\vspace{1.0em}
\plotone{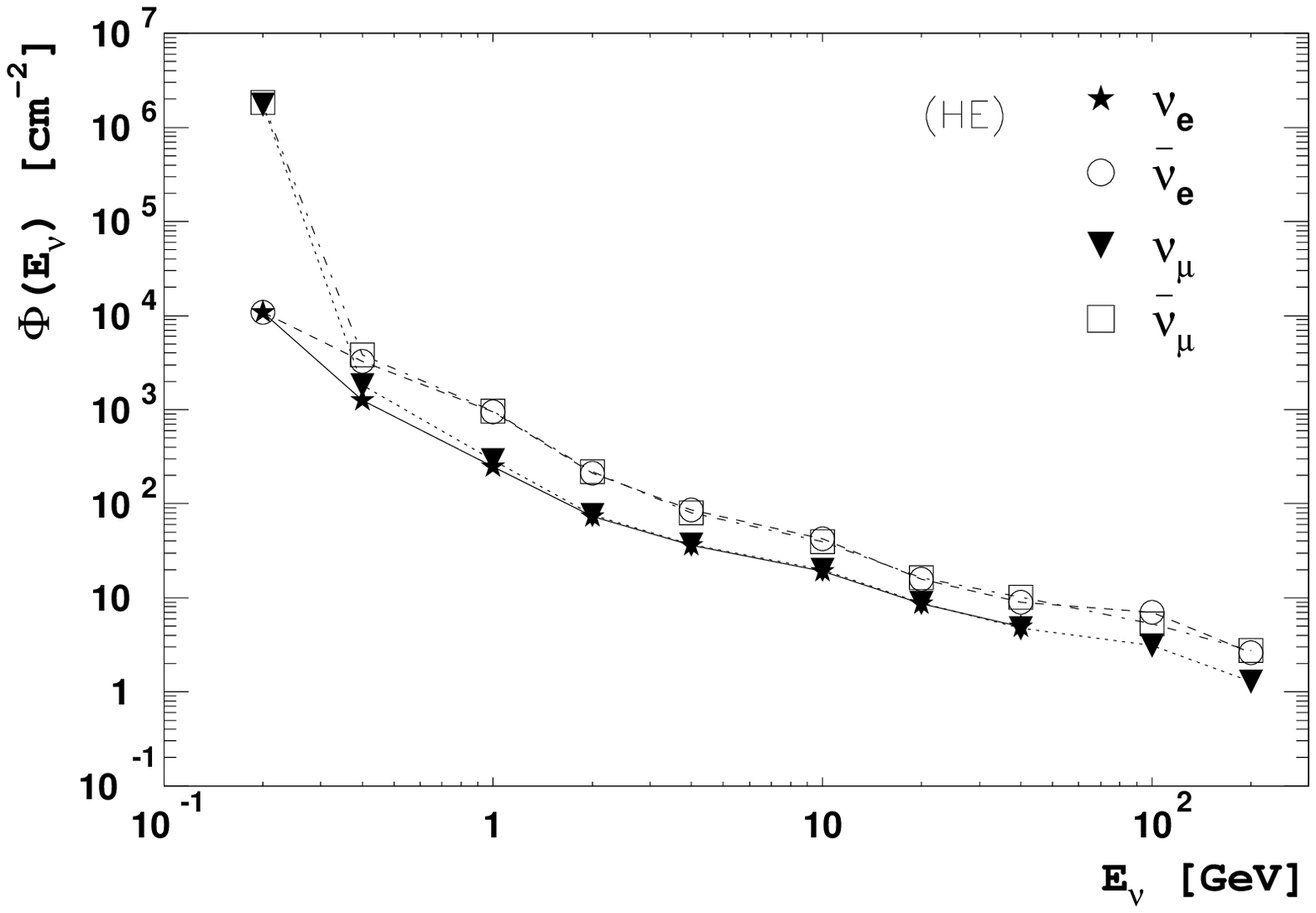}\vspace{1.0em}
\plotone{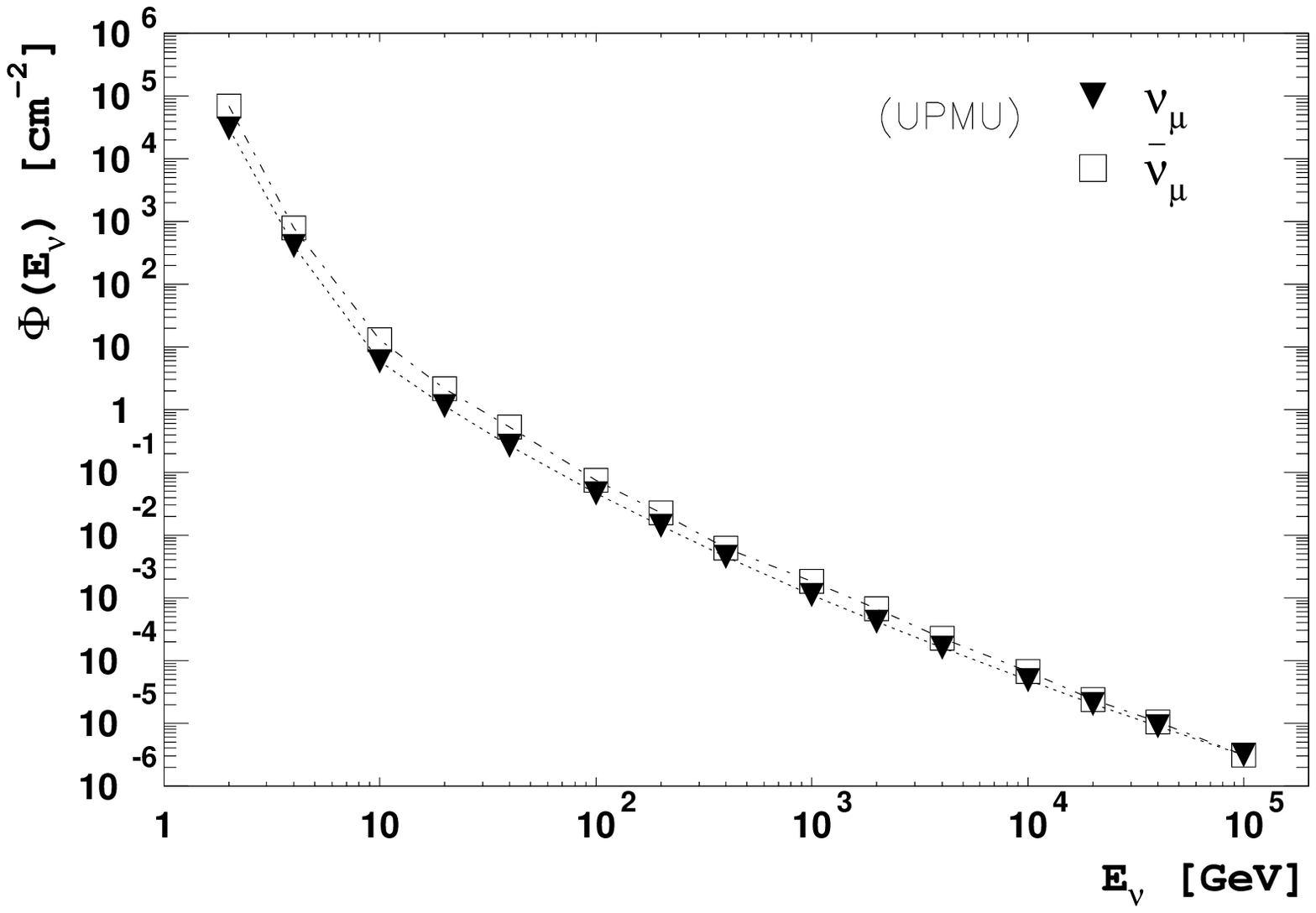}
\caption{\label{f:nu_flux} \small The \gf, $\Phi(E_\nu)$, of 90\% C.L. upper limits on GRB neutrino and antineutrino fluence per GRB obtained from the LE, HE, and upmu data samples, respectively.  This \gf\, represents the fluence upper limit for monoenergetic neutrinos and antineutrinos at different energies.  To obtain the physical fluence limit, one needs to convolute $\Phi(E_\nu)$ with a neutrino energy spectrum.  The fluence limit for $\nu_e$ ends at $40\,$GeV, because of \Sk's limited sensitivity to $\nu_e$ at the highest energies.}
\end{figure}

Figure~\ref{f:nu_flux} shows the fluence limit \gf, $\Phi(E_\nu)$, for $\nu_e$, $\bar{\nu}_e$, $\nu_\mu$, and $\bar{\nu}_\mu$ obtained from \Sk's LE, HE, and upmu data samples.  The calculated values of $\Phi$ at each neutrino energy for the three analyses are shown in Tables~\ref{t:le_flux}, \ref{t:he_flux}, and \ref{t:upmu_flux}, respectively.  In the calculation we assume that all the neutrinos are emitted with the same flavor.  Note that Table~\ref{t:le_flux} (energies above $100\,$MeV) includes $\Phi_{\nu_\mu}$ and $\Phi_{\bar{\nu}_\mu}$ coming from possible decay electron interactions from ``stealth muons'', i.e. muons or antimuons created in the $105-140\,$MeV range (below the Cherenkov threshold) by quasi-elastic scattering of $\nu_\mu$ and $\bar{\nu}_\mu$.  These limits are up to a factor of $10^3$ lower than the limits for the $7-80\,$MeV range.  This increase in sensitivity comes from the quasi-elastic scattering cross section that becomes dominant for $\nu_\mu$ and $\bar{\nu}_\mu$ in the $105-140\,$MeV range.

In order to demonstrate \Sk's sensitivity to GRB neutrinos, we assumed an $E^{-2}$ neutrino spectrum to calculate the 90\% C.L. upper limit on $\nu_e$, $\bar{\nu}_e$, $\nu_\mu$, and $\bar{\nu}_\mu$ total fluence.  Table~\ref{t:F} shows this total fluence calculated for neutrinos produced in \Sk's three neutrino energy ranges: $\rm 7\,MeV-80\,MeV$ (from LE sample), $\rm 200\,MeV-200\,GeV$ (from HE sample), and $\rm 2\,GeV-100\,TeV$ (from upmu sample).  Note that for $\nu_\mu$ and $\bar{\nu}_\mu$ the fluence limit in the LE range does not include the ``stealth muon'' limits ($105-140\,$MeV range).  If we extended our LE neutrino energy range up to $\sim140\,$MeV, our fluence upper limits for $\nu_\mu$ and $\bar{\nu}_\mu$ in Table~\ref{t:F} would become $\rm 2.64\times10^{6}\,cm^{-2}$ and $\rm 3.48\times10^{6}\,cm^{-2}$, respectively, which is better than the existing limit roughly by a factor of $10^2$.  

For comparison, we calculated a rough estimate of the expected neutrino fluence, $F$,  from a cosmological GRB.  In the calculation we assumed an $E^{-2}$ neutrino spectrum, a distance to the GRB of z=1 ($\rm D_L\approx6.6\,$Gpc), the total neutrino energy of $10^{53}\,$erg \cite{r:kumar} emitted isotropically, the fact that all emitted neutrinos have the same flavor, and that the energy range of emitted neutrinos coincides with the energy range of each data sample.  Using these assumptions, we predict a neutrino fluence from a single GRB to be $\rm 1.4\,cm^{-2}$ (LE), $\rm 1.7\times10^{-2}\,cm^{-2}$ (HE), $\rm 1.1\times10^{-3}\,cm^{-2}$ (upmu), which is roughly a factor of $\sim10^6$, $10^4$, and $30$ lower than our best upper limit in the three energy ranges, respectively.

Comparing \Sk's fluence limits to previously published GRB neutrino fluence upper limits from the IMB detector \cite{r:imb}, we find that our limits are better roughly by a factor of $10^2$ for energies $\geq$$1\,$GeV.  
Our fluence limit on upward-going muons per average burst ($\rm 0.31\times10^{-9}\,cm^{-2}$) is comparable to a similar limit by the MACRO detector \cite{r:macro}.
Our neutrino fluence limits are consistent with the recent predictions for \sk of a fireball model with a collapsar progenitor \cite{r:sato}.

\section{Conclusion}

A time and direction correlation analysis of \SK\, events with BATSE gamma-ray bursts was performed for \Sk's solar neutrino, atmospheric neutrino, and upward-going muon data samples.  No signal in excess of the expected background fluctuations was found.  A 90\% C.L. fluence upper limits on neutrino and antineutrino emission from an average GRB were calculated at various neutrino energies detected in \Sk.  We also calculated the total GRB neutrino fluence upper limits for an $E^{-2}$ spectrum in \Sk's three energy ranges.  We found that our limits are {\it at least} a factor of $30$ higher than our rough estimate of GRB neutrino emission and are consistent with most model predictions.  \Sk's limits are the most stringent fluence upper limits for GRB neutrinos at energies $\rm 7\,MeV-100\,TeV$.

\acknowledgements

The authors acknowledge the cooperation of the Kamioka Mining and
Smelting Company.  The Super-Kamiokande detector has been built and
operated from funding by the Japanese Ministry of Education, Culture,
Sports, Science and Technology, the U.S. Department of Energy, and the
U.S. National Science Foundation, with support for individual researchers 
from Research Corporation's Cottrell College Science Award.

\newpage


\newpage


\begin{deluxetable}{ccccccc} 
\tabletypesize{\small}
\tablecolumns{8} 
\tablewidth{0pc} 
\tablecaption{\label{t:results} Summary of the maximum signal bin of the correlation analyses.}
\tablehead{
\colhead{} & \colhead{} & \colhead{} & \colhead{} & \colhead{} & \colhead{Number of GRBs} \\
\colhead{Window} & \colhead{Sample} & \colhead{GRBs} & 
\colhead{Bgd.$^{\rm{a}}$} & \colhead{Sig.$^{\rm{b}}$} & 
\colhead{with Max. Sig.$^{\rm{c}}$} & \colhead{Prob.$^{\rm{d}}$}
}
\startdata
\vspace{-0.2em}
$\pm10\,$s   & (LE) & 1086 & 0.0160 & 1 & (11 GRBs) & 0.96  \\ \vspace{0.2em}
             & (HE) & 1115 & 0.00198& 1 & (3 GRBs) & 0.38 \\ \vspace{-0.2em}
$\pm100\,$s  & (LE) & 1081 & 0.160  & 4 & GRB 970605 & 0.026 \\ \vspace{0.2em}
             & (HE) & 1111 & 0.0198 & 1 & (26 GRBs) & 0.21  \\ \vspace{-0.2em}
$\pm1000\,$s & (LE) & 1027 & 1.60   & 7 & GRB 991004D, NTB 961019.51 & 0.40 \\ \vspace{0.2em}
             & (HE) & 1056 & 0.198  & 2 & (19 GRBs) & 0.45  \\ \vspace{-0.2em}
\vspace{-1em} \\ \cline{1-7} \vspace{-0.8em} \\
SN-like: ($-8\,$hr,$-7\,$hr) & (LE) & 1018 & 2.87  &11 & GRB 980601  & 0.99 \\ \vspace{0.2em}
                        & (HE) & 1037 & 0.357 & 5 & GRB 990202B & 0.59 \\ \vspace{-0.2em}
AG: ($+4\,$hr,$+5\,$hr) & (LE) & 990  & 2.87  & 9 & NTB 970525.36 & $\sim$1.00 \\ \vspace{0.2em}
                        & (HE) & 1011 & 0.357 & 5 & GRB 990202A & 0.58 \\ \vspace{-0.2em}
\vspace{-1em} \\ \cline{1-7} \vspace{-0.8em} \\
$\pm1000\,$sec; $\le$$15^\circ$ & (upmu) & 1454 & 0.00096 & 1 & GRB 991004D & 0.75 \\ 
\enddata
\tablenotetext{a}{\small Expected background for the corresponding search time window for a single GRB.}
\tablenotetext{b}{\small Maximum signal (maximum number of observed events) for a single GRB.}
\tablenotetext{c}{\small Number of GRBs with maximum signal. For one or two GRBs, the BATSE current catalog number is given \cite{r:batse,r:kommers}.}
\tablenotetext{d}{\small Probability to get at least the observed number of GRB candidates with maximum signal, given the expected background and the total number of GRBs analyzed.  The probabilities for the SN-like and AG search window representatives include the trials factor for the number of search windows used (24).}
\end{deluxetable} 

\begin{deluxetable}{cccccc}
\tabletypesize{\small}
\tablecolumns{6} 
\tablewidth{0pc} 
\tablecaption{\label{t:number} The 90\% C.L. upper limit on number of \sk events per GRB, $N_{90}$.}
\vspace{0.2cm}
\tablehead{
\colhead{Sample} & \colhead{GRBs} & \colhead{Search Window} &
\colhead{Tot. Bgd.$^{\rm{a}}$} & \colhead{Tot. Sig.$^{\rm{b}}$} & 
\colhead{N$_{90}$}
}
\startdata
LE & 1081 & $\pm100\,$s & 173 & 177 & $24.1\times10^{-3}$ \\
HE e-like & 1111 & $\pm100\,$s & 12.3  & 16  & $9.5\times10^{-3}$ \\
HE $\mu$-like & 1111 & $\pm100\,$s & 9.7  & 14  & $9.6\times10^{-3}$ \\
upmu & 1454 & $\pm1000\,$s, $\le$$15^\circ$ & 0.67 & 1 & $2.35\times10^{-3}$ \\
\enddata
\tablenotetext{a}{\small Total, cumulated expected background for all the GRBs.}
\tablenotetext{b}{\small Total, cumulated number of observed events for all the GRBs.}
\end{deluxetable} 

\begin{deluxetable}{rccccc}
\tabletypesize{\small}
\tablecolumns{6} 
\tablewidth{0pc} 
\tablecaption{\label{t:le_flux} GRB neutrino fluence \gf\, upper limits ($90\%$~C.L.) obtained from the LE neutrino data sample.}
\vspace{0.2cm}
\tablehead{
\colhead{$\rm E_\nu$} & &
\colhead{$\rm \Phi_{\nu_e}$\small$\rm[cm^{-2}]$} & 
\colhead{$\rm \Phi_{\bar{\nu}_e}$\small$\rm[cm^{-2}]$} &
\colhead{$\rm \Phi_{\nu_\mu}^{\rm a}$\small$\rm[cm^{-2}]$} & 
\colhead{$\rm \Phi_{\bar{\nu}_\mu}^{\rm a}$\small$\rm[cm^{-2}]$}
}
\startdata
7 MeV && 1.05$\times10^{9}$ & 1.47$\times10^{8}$ & 6.41$\times10^{9}$ & 6.41$\times10^{9}$ \\
10 MeV && 1.53$\times10^{8}$ & 4.67$\times10^{6}$ & 9.37$\times10^{8}$ & 9.37$\times10^{8}$ \\
14 MeV && 6.79$\times10^{7}$ & 1.90$\times10^{6}$ & 4.11$\times10^{8}$ & 4.11$\times10^{8}$ \\
20 MeV && 3.84$\times10^{7}$ & 9.09$\times10^{5}$ & 2.30$\times10^{8}$ & 2.30$\times10^{8}$ \\
30 MeV && 2.28$\times10^{7}$ & 4.50$\times10^{5}$ & 1.37$\times10^{8}$ & 1.37$\times10^{8}$ \\
50 MeV && 1.35$\times10^{7}$ & 2.33$\times10^{5}$ & 7.94$\times10^{7}$ & 7.94$\times10^{7}$ \\
80 MeV && 9.14$\times10^{6}$ & 1.84$\times10^{5}$ & 5.28$\times10^{7}$ & 5.28$\times10^{7}$ \\
\cline{1-6}
105 MeV && \nodata & \nodata & 7.55$\times10^{4}$ & 1.43$\times10^{5}$ \\
120 MeV && \nodata & \nodata & 5.56$\times10^{4}$ & 6.09$\times10^{4}$ \\
140 MeV && \nodata & \nodata & 2.33$\times10^{4}$ & 3.17$\times10^{4}$ \\
\enddata
\tablenotetext{a}{\small The values of $\Phi_{\nu_\mu}$ and $\Phi_{\bar{\nu}_\mu}$ above $105\,$MeV come from ``stealth muons'' produced by neutrinos in the energy range $105-140\,$MeV (see \S\ref{s:flux} for details).}
\end{deluxetable} 

\begin{deluxetable}{rccccc}
\tabletypesize{\small}
\tablecolumns{6} 
\tablewidth{0pc} 
\tablecaption{\label{t:he_flux} GRB neutrino fluence \gf\, upper limits ($90\%$~C.L.) obtained from the HE neutrino data sample.}
\vspace{0.2cm}
\tablehead{
\colhead{$\rm E_\nu$} & &
\colhead{$\rm \Phi_{\nu_e}$\small$\rm[cm^{-2}]$} & 
\colhead{$\rm \Phi_{\bar{\nu}_e}$\small$\rm[cm^{-2}]$} &
\colhead{$\rm \Phi_{\nu_\mu}$\small$\rm[cm^{-2}]$} & 
\colhead{$\rm \Phi_{\bar{\nu}_\mu}$\small$\rm[cm^{-2}]$}
}
\startdata
0.2 GeV && 1.08$\times10^{4}$ & 1.08$\times10^{4}$ & 1.74$\times10^{6}$ & 1.84$\times10^{6}$ \\
0.4 GeV && 1.25$\times10^{3}$ & 3.27$\times10^{3}$ & 1.83$\times10^{3}$ & 3.84$\times10^{3}$ \\
1 GeV && 2.49$\times10^{2}$ & 9.52$\times10^{2}$ & 2.88$\times10^{2}$ & 9.58$\times10^{2}$ \\
2 GeV && 7.28$\times10^{1}$ & 2.11$\times10^{2}$ & 7.54$\times10^{1}$ & 2.17$\times10^{2}$ \\
4 GeV && 3.61$\times10^{1}$ & 8.58$\times10^{1}$ & 3.67$\times10^{1}$ & 8.02$\times10^{1}$ \\
10 GeV && 1.91$\times10^{1}$ & 4.21$\times10^{1}$ & 1.99$\times10^{1}$ & 3.95$\times10^{1}$ \\
20 GeV && 8.62$\times10^{0}$ & 1.57$\times10^{1}$ & 8.85$\times10^{0}$ & 1.63$\times10^{1}$ \\
40 GeV && 4.90$\times10^{0}$ & 8.92$\times10^{0}$ & 4.76$\times10^{0}$ & 1.02$\times10^{1}$ \\
100 GeV && \nodata & 6.98$\times10^{0}$ & 3.13$\times10^{0}$ & 5.31$\times10^{0}$ \\
200 GeV && \nodata & 2.61$\times10^{0}$ & 1.28$\times10^{0}$ & 2.73$\times10^{0}$ \\
\enddata
\end{deluxetable} 

\begin{deluxetable}{rccc}
\tabletypesize{\small}
\tablecolumns{6} 
\tablewidth{0pc} 
\tablecaption{\label{t:upmu_flux} GRB neutrino fluence \gf\, upper limits ($90\%$~C.L.) obtained from the upmu data sample.}
\vspace{0.2cm}
\tablehead{
\colhead{$\rm E_\nu$} & &
\colhead{$\rm \Phi_{\nu_\mu}$\small$\rm[cm^{-2}]$} & 
\colhead{$\rm \Phi_{\bar{\nu}_\mu}$\small$\rm[cm^{-2}]$}
}
\startdata
2 GeV && 3.08$\times10^{4}$ & 6.98$\times10^{4}$ \\
4 GeV && 4.12$\times10^{2}$ & 7.96$\times10^{2}$ \\
10 GeV && 5.87$\times10^{0}$ & 1.32$\times10^{1}$ \\
20 GeV && 1.13$\times10^{0}$ & 2.18$\times10^{0}$ \\
40 GeV && 2.63$\times10^{-1}$ & 5.22$\times10^{-1}$ \\
100 GeV && 4.62$\times10^{-2}$ & 7.53$\times10^{-2}$ \\
200 GeV && 1.39$\times10^{-2}$ & 2.25$\times10^{-2}$ \\
400 GeV && 4.50$\times10^{-3}$ & 6.22$\times10^{-3}$ \\
1 TeV && 1.12$\times10^{-3}$ & 1.83$\times10^{-3}$ \\
2 TeV && 4.15$\times10^{-4}$ & 6.77$\times10^{-4}$ \\
4 TeV && 1.61$\times10^{-4}$ & 2.23$\times10^{-4}$ \\
10 TeV && 4.87$\times10^{-5}$ & 6.72$\times10^{-5}$ \\
20 TeV && 2.05$\times10^{-5}$ & 2.41$\times10^{-5}$ \\
40 TeV && 8.93$\times10^{-6}$ & 1.04$\times10^{-5}$ \\
100 TeV && 3.08$\times10^{-6}$ & 3.08$\times10^{-6}$ \\
\enddata
\end{deluxetable}

\begin{deluxetable}{rcccccc}
\tabletypesize{\small}
\tablecolumns{6} 
\tablewidth{0pc} 
\tablecaption{\label{t:F} GRB neutrino total fluence upper limits ($90\%$~C.L.) for a $E^{-2}$ spectrum in \Sk's three energy ranges.}
\vspace{0.2cm}
\tablehead{
\colhead{$\rm E_\nu\; range$} & &
\colhead{$\rm F_{\nu_e}$\small$\rm[cm^{-2}]$} & 
\colhead{$\rm F_{\bar{\nu}_e}$\small$\rm[cm^{-2}]$} &
\colhead{$\rm F_{\nu_\mu}$\small$\rm[cm^{-2}]$} & 
\colhead{$\rm F_{\bar{\nu}_\mu}$\small$\rm[cm^{-2}]$} &
\colhead{$\rm Prediction^{\rm a}$\small$\rm[cm^{-2}]$}
}
\startdata
7 MeV - 80 MeV && 4.44$\times10^{7}$ & 9.52$\times10^{5}$ & 2.65$\times10^{8}(^{\rm b})$ & 2.65$\times10^{8}(^{\rm b})$ & 1.4 \\
0.2 GeV - 200 GeV && 1.66$\times10^{2}$ & 2.97$\times10^{2}$ & 1.39$\times10^{2}$ & 3.00$\times10^{2}$ & 1.7$\times10^{-2}$ \\
2 GeV - 100 TeV && \nodata & \nodata & 3.83$\times10^{-2}$ & 4.96$\times10^{-2}$ & $1.1\times10^{-3}$ \\
\enddata
\tablenotetext{a}{\small The prediction of a GRB neutrino fluence, assuming $E^{-2}$ spectrum and the matching energy range, z=1 distance, and $10^{53}$ total energy emitted in neutrinos.}
\tablenotetext{b}{\small These limits become $\rm 2.64\times10^{6}\,cm^{-2}$ and $\rm 3.48\times10^{6}\,cm^{-2}$, respectively, if we extend the LE energy region to $\sim140\,$MeV, in order to include neutrinos resulting in ``stealth muons'' (see \S\ref{s:flux} for details).}
\end{deluxetable}


\end{document}